\documentclass[acmtog,nonacm]{acmart}
\acmSubmissionID{480}
\AtBeginDocument{%
  \providecommand\BibTeX{{%
    \normalfont B\kern-0.5em{\scshape i\kern-0.25em b}\kern-0.8em\TeX}}}

\usepackage{diagbox}
\usepackage{amsfonts}

\usepackage{amsmath}
\DeclareMathOperator*{\argmin}{argmin} 

\def\RR{\mathbb{R}}

\newcommand{\hide}[1]{}

\def\shortcite{\cite}

\begin{document}

\title{A Rotation-Strain Method to Model Surfaces using Plasticity}

\author{Jiahao Wen}
\affiliation{%
  \institution{University of Southern California}
  \city{Los Angeles}
  \country{USA}}
\email{jiahaow@usc.edu}

\author{Bohan Wang}
\affiliation{%
 \institution{University of Southern California, Massachusetts Institute of Technology}
 \city{Boston}
 \country{USA}}
\email{bohanwan@usc.edu}

\author{Jernej Barbi\v{c}}
\affiliation{%
  \institution{University of Southern California}
  \city{Los Angeles}
  \country{USA}}
\email{jnb@usc.edu}


\begin{abstract}
Modeling arbitrarily large deformations of surfaces smoothly embedded in three-dimensional space is challenging. The difficulties come from two aspects: the existing geometry processing or forward simulation methods penalize the difference between the current status and the rest configuration to maintain the initial shape, which will lead to sharp spikes or wiggles for large deformations; the co-dimensional nature of the problem makes it more complicated because the deformed surface has to locally satisfy compatibility conditions on fundamental forms to guarantee a feasible solution exists. To address these two challenges, we propose a rotation-strain method to modify the fundamental forms in a compatible way, and model the large deformation of surface meshes smoothly using plasticity.  The user prescribes the positions of a few vertices, and our method finds a smooth strain and rotation field under which the surface meets the target positions. We demonstrate several examples whereby triangle meshes are smoothly deformed to large strains while meeting user constraints.


\end{abstract}




\begin{teaserfigure}
\begin{centering}
  \includegraphics[width=1.0\textwidth]{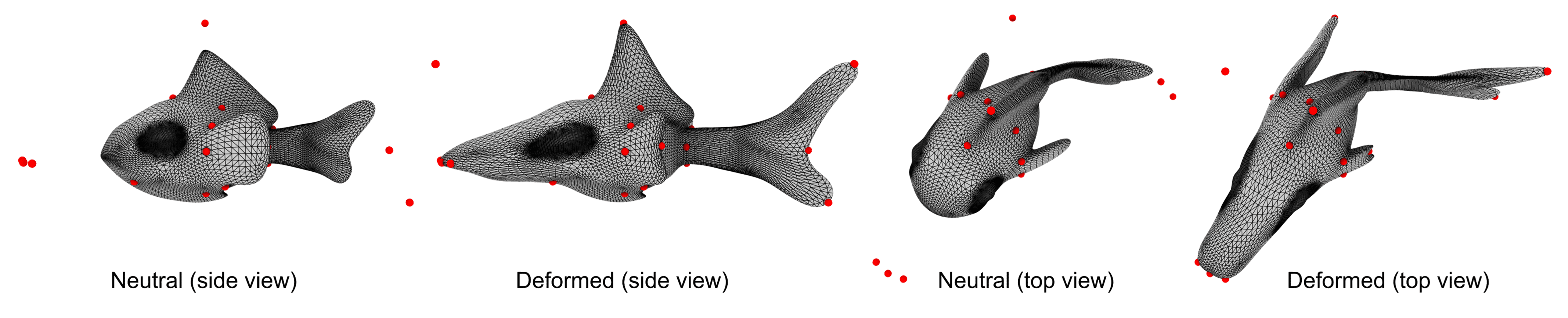}
  \vspace{-0.75cm}
  \caption 
  {\textbf{Large strain editing of a fish:}
   We apply landmarks (red dots) to edit the fish. Our method scales the fish non-uniformly. We stretch the mouth and a side fin of the fish, 
   exaggerate the tail and shrink the other side-fin at the same time.  Our methods can successfully produce a smooth result.} 
  \label{fig:teaser}
  \end{centering}
\end{teaserfigure}

\maketitle

\section{Introduction}
\label{sec:intro}

\hide{
Must have figures:

Fish [4 views; teaser 1x4]
Plane [comparison to hinge-based energy]
Deforming a muscle to match medical (MRI) landmarks
Horse [2x2 figure; same layout as now; replace with new results; drop the tail landmark]
Keep Figure 8 as is
Bunny Fig 2 (comparison to ShapeOP; drop ARAP; make figure into a single-column)
Benchmark [2x2 figure]
Sphere
Torus [incorporate 2012 result]
Arch

Good to have figures:

Abandon figures:

Large strain editing of the hand [replace ARAP and ShapeOp with 
new results; highlight problem areas with rectangles]
Maple leaf (previous teaser)

Video: 
Fish, hand, arch: compare to ARAP, ShapeOP, Primo
Fish: show optimization progress
}

Modeling surfaces and their deformations is a central topic in computer graphics. 
In the context of surface modeling, a surface representing the shape needs to be stretched, 
sheared, or rotated to meet arbitrary user constraints. 
Users intuitively expect that the deformation should always preserve the intrinsic original 
shape of the object, while being  
globally smooth. 
To achieve this goal, standard geometric shape modeling methods and physically-based simulation methods 
penalize the changes between the current status and the rest configuration in a smooth manner to maintain the origin shape.
Given point constraints, however, these standard methods either produce spikes or excessive curvatures under large deformation.
To address this, plastic deformation can be used because it changes the rest shape directly. 
Wang et al.~\shortcite{Wang:2021:MOP} used plastic strains to deform volumetric meshes of template human organs to match medical images of a real subjects,  producing smooth results satisfying sparse position constraints. 
At a first glance, it seems that applying such methodology---i.e.,the plastic deformation method---directly onto surface is straightforward. 
However, this is not the case and we notice large difficulties therein.
When we talk about the surface mesh deformation, 
we are referring to embedding a 2D manifold into 3D space, 
which is a co-dimensional problem compared to a 3D volumetric mesh. 
Therefore, the plasticity Degrees of Freedom (DoFs) 
contain intrinsic (in-plane) and extrinsic parts (bending).
Moreover, given some arbitrarily defined local plasticity, 
we must ensure that such a surface even exists (at least locally), so that we can reconstruct the mesh vertex positions by solving for a static equilibrium
under the given plastic strains.
In other words, the definition of intrinsic and extrinsic plasticity must satisfy compatibility conditions on fundamental forms.  
To achieve these requirements, we propose a novel differential representation of surface shape deformation,
which models the local surface deformation using a $3\times 3$ rotation and
a $2\times 2$ symmetric matrix. 
The rotation globally orients every local surface ``patch'' in the world, 
and the symmetric matrix models an arbitrarily oriented (large) 
strain of the surface. This new representation is very versatile
and can easily model large spatially varying surface strains.

The representation can be seen as an extension of rotation-strain coordinates~\cite{Huang:2011:ISI} 
from solids in $\RR^3$ to surfaces embedded into $\RR^3.$
It automatically makes the parameterization gradient locally integrable, 
i.e., the surface locally satisfies the Gauss-Codazzi equations~\cite{Chen:2018:PSO} 
and therefore locally exists.
This greatly improves the convergence of shape optimization.
To use our representation to perform geometric shape modeling, 
we define an optimization problem that optimizes 
for smooth spatially varying $3\times 3$ rotations $R$
and $2\times 2$ symmetric matrices $S$ so that the resulting surface matches the user constraints as closely as possible.
Our optimization successfully handles difficult yet sparse point constraints in a few different examples 
and produces smooth output shapes under large and anisotropic shape deformation.
We demonstrate that these effects are difficult to achieve
with previous methods.
Our contributions are mostly two-fold:
\begin{itemize}
    \item To the best of our knowledge, we are first to use plasticity and FEM to successfully model surface meshes undergoing large strains and rotations, specified by point constraints on the boundary of the object. Through plasticity, we can generate results that are impossible to obtain using elastic deformation, such as bending and inflating an object. Compared to previous works which mainly penalize the deformation magnitude, our method additionally defines "smoothness" energy on plasticity DoFs so that we can generate smooth results. By experiments, our method succeeds in the extreme benchmarks and produces expected results that previous methods cannot achieve.
    \item We propose a new Rotation-Strain method to modify the first and second fundamental forms of a (discretized) surfaces embedded into the 3-dimensional space. Compared to  \cite{Chen:2018:PSO} and \cite{Kircher:2008:FFM}, our method can cleanly decouple the intrinsic and extrinsic DoFs, thus guaranteeing compatibility for each local patch (triangle in discrete configuration), which is proved to be crucial for decreasing the objective energy.
\end{itemize}

\section{Related Work}
\label{sec:relatedWork}
\begin{figure}[!t]
            \includegraphics[width=1.0\hsize]{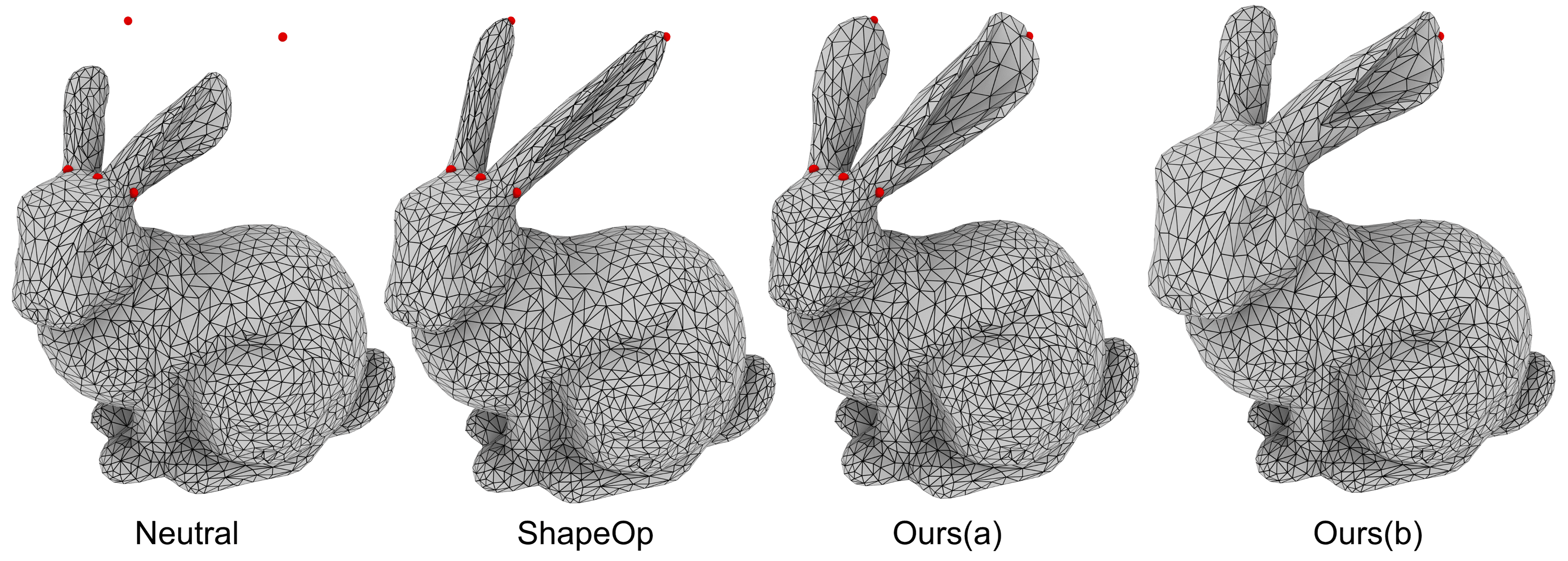}
        \caption[Our method scales the object in a manner that preserves the shape.]
        {\textbf{Our method scales the object in a manner that preserves the shape:}
        Observe that ShapeOp~\cite{Bouaziz:2014:PD}
        produce suboptimal results, under identical inputs as our method (a):
        the bunny has fixed landmarks (red dots) at the bunny's bottom and 
        the base of the ears, and a 
        landmark on each ear which is positioned to command the ear to stretch.
        Tweaking the parameters of ShapeOp did not improve the result.
        In case (a), observe how the stretched ear grows to meet the landmark,
        while preserving its inherent geometric shape (the ears are still curled). In case (b), we removed
        the landmarks at the base of the ears and kept only a single ear landmarked. 
        As expected, the single landmark
        now causes the entire bunny to uniformly grow in size.
        }
        \label{fig:bunny}
\end{figure}
Modeling shape deformation is an important topic in computer graphics 
and encountered in many sub-disciplines, including geometric modeling~\cite{Alexa:2006:ISM,Botsch:2008:OLV}, 
physically based FEM simulation~\cite{Sifakis:2015:FEM} and mesh non-rigid registration~\cite{Allen:2003:TSO}.
In general, a shape deformation problem can be treated as
an optimization problem whose objective function is
defined as the ``smoothness'' of the shape, combined with
some user constraints.

Different definitions of the ``smoothness'' energies give
rise to various output properties.
The smoothness of the shape has been formulated 
through variational methods~\cite{Botsch:2004:AIF,Wang:2015:LSD},
Laplacian surface editing~\cite{Sorkine:2004:LSE}, 
as-rigid-as-possible (ARAP) deformation~\cite{Igarashi:2005:ARA,Sorkine:2007:ARA}, 
coupled prisms (PRIMO)~\cite{Botsch:2006:PCP},
and partition-of-unity interpolation weights such as 
bounded biharmonic weights (BBW)~\cite{Jacobson:2011:BBW}
and quasi-harmonic weights~\cite{Wang:2021:FQW}.
Nonetheless, existing methods produce suboptimal deformations 
when the shape undergoes large rotations and large strains simultaneously (Figure~\ref{fig:bunny}),
or they are not designed for meeting user vertex constraints. 
Variational methods suffer from large rotations,
while ARAP and PRIMO produce spiky shapes given difficult constraints.
These problems have been discussed and illustrated by~\cite{Botsch:2008:OLV,Wang:2021:MOP}.
BBW generates smooth interpolation weights that can be used for interpolating transformations across the entire shape, 
but it takes a substantial amount of time to compute the weights.
In addition, to generate interpolation weights, the handles must be predefined.
To overcome the problem caused by large rotations in variational methods,
the smoothness energy has been defined to penalize the Laplacian after the rotations~\cite{Bouaziz:2012:ShapeUp,Bouaziz:2014:PD}.
While the resulting shapes are C2-smooth, 
the method generates wiggly artifacts similar to variational methods~\cite{Wang:2021:FQW}.
To address this problem and handle large strains properly,
the rest shape can be reset during each deformation iteration~\cite{Gilles:2006:AMO,Schmid:2009:MSM,Gilles:2010:MMS}.
Doing so, however, loses important features of the original input shape,
such as not being able to preserve the volume and the sharp features.
In contrast, our method can solve these problems and handle large rotations and large strains correctly.

When performing non-rigid registration on a template mesh to match the target shape,
a smoothness energy is often needed in the objective function.
Existing methods penalize the affine transformations between neighboring vertices
or triangles~\cite{Allen:2003:TSO,Amberg:2007:OSN,Li:2008:GCO}.
Such energy is combined with dense correspondences to  
achieve deforming the template mesh to match the target.
When the smoothness energy is combined with sparse inputs, however, 
artifacts similar to those in variational methods may appear, as demonstrated by~\cite{Wang:2021:MOP}.
Deformation transfer is able to handle sparse markers without visible artefacts~\cite{Sumner:2004:DTF}, but it cannot handle large rotations.
Kircher et al. solved the problem by decomposing the affine transformations into the rotational and shear/stretch parts,
and treated them separately~\cite{Kircher:2008:FFM}.

Physically-based FEM simulation can also be used to deform the shape.
An object can either be treated as volumetric object or surface object.
For volumetric objects, the elastic energy penalizes volume changes, large strains,
and gives C0 continuity around constraints~\cite{Sifakis:2015:FEM}, similar to ARAP.
For surface objects, they are modeled using cloth simulation.
The cloth elastic energy is decoupled into two terms, namely,
in-plane elastic energy and bending energy.
The in-plane elastic energy is defined in a similar fashion to the elastic energy for 3D volumetric objects~\cite{Volino:2009:ASA},
and, as a result, it shares the same artifacts as volumetric simulation.
On the other hand, the bending energy is modeled as the change of curvature.
In fact, penalizing the bi-Laplacian mentioned above in variational methods 
can also be considered as a type of a bending energy, as it models the change of mean curvature.
For a comprehensive comparison of bending models, we refer readers to~\cite{Chen:2018:PSO}.
In general, elastic models are always large-strain-unfriendly, because they penalize the changes to the rest shape.

Plastic deformation can be used to solve such problems, because 
it models shapes that undergo permanent and large deformation
and therefore it is in principle a natural choice to do shape modeling.
However, plastic deformation is only widely used in forward simulation~\cite{Obrien:2002:GMA,Mueller:2004:IVM,Irving:2004:IFE,Bargteil:2007:FEM,Stomakhin:2013:AMP,Chen:2018:PBF}.
Wang et al.~\shortcite{Wang:2021:MOP} introduce plastic deformation to shape modeling,
but they only addressed volumetric objects (tetrahedral meshes in 3D), which limits the applications.
In our work, we address surfaces in 3D. Our definition of DoFs makes it
possible to handle both rotations and anisotropic stretching, while their method only permits symmetric $3\times 3$ plastic transformations, i.e.,
(anisotropic) stretching only. Because we formulate the
plastic deformations on the surface object, our method uses fewer DoFs, which leads to faster performance. Last but most important, presence of the co-dimension makes the problem substantially more difficult, due to the necessity to model compatible first and second fundamental forms. \cite{lipman2005linear} proposed a rotation-invariant representation of surface meshes using discrete frames and fundamental form coefficients. It decouples frame rotation from the deformation of the tangent space and solves them in two separate steps. However, it doesn't guarantee the compatibility of fundamental forms. \cite{wang2012linear} used edge lengths and dihedral angles as primary variables to compute discrete fundamental forms, and derived local and global compatibility conditions. However, their method penalizes the distance between the current shape to the compatible initial shape in the  reconstruction of the surface similarly, and will still suffer from sharp spikes under large deformation (Figure \ref{fig:arch}).  \cite{chern2018shape} discusses the isometric immersion problem, taking as input an orientable surface triangle mesh annotated with edge lengths only and outputting vertex positions. \cite{martinez2014smoothed} penalizes the magnitude of energy gradients over the whole mesh and achieves smooth shapes compared to ARAP, however, as discussed in \cite{Wang:2021:MOP}, such methods suffer from wiggles under large deformations. Although some previous methods discussed different ways to discretize fundamental forms and derive the compatibility of surface meshes, it is very hard to define plasticity and the corresponding smoothness based on their definitions. 
That is why we choose to follow \cite{Chen:2018:PSO}.
  
Parameterizing plastic deformation on a surface is not straightforward: 
an incorrect choice of the model will produce degenerate outputs and non-convergence of optimization.
To achieve our goal, we first choose a state-of-the-art thin-shell simulation model~\cite{Chen:2018:PSO}.
It not only defines a physically-accurate simulation, but also 
gives a proper space to define plastic deformations.
We also experimented with other elastic cloth models 
such as~\cite{Grinspun:2003:DS}, 
but the plastic deformation is limited by the elastic model 
and causes artefacts (Figure~\ref{fig:plane}).
Methods that define the plastic deformation on thin-shells are rare.
We first attempted to use the plastic model defined in~\cite{Chen:2018:PSO},
but the optimization failed to converge even on a simple input.
Directly decomposing a plastic deformation gradient into a rotation and a symmetric matrix is also not doable in practice~\cite{Kircher:2008:FFM} (Section~\ref{sec:continuousFormulation}). 

\section{Plasticity of Surfaces}
\label{sec:shell}

Before describing our method for shape modeling of surfaces, 
we give a brief review of the surface theory in differential geometry,
following~\cite{Chen:2018:PSO}.

\subsection{Continuous Formulation}
\label{sec:continuousFormulation}

\begin{figure}[!t]
            \includegraphics[width=1.0\hsize]{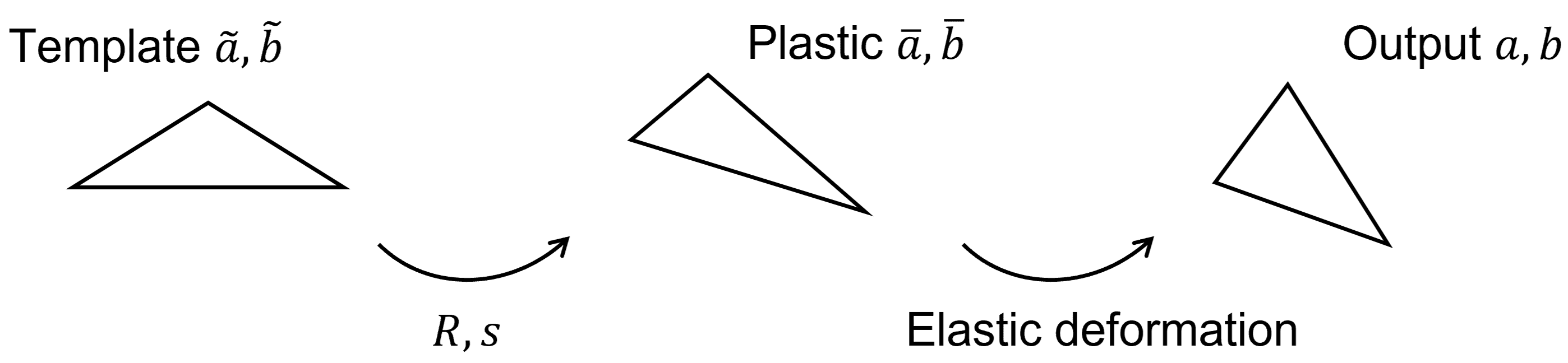}
        \caption
        {\textbf{Plastic and elastic deformation of a single triangle.}
        }
        \label{fig:plastic_schematic}
\end{figure}

\paragraph{Thin-Shell Representation} 

Our modeling of surfaces stems from the continuous mechanics where a
surface is modeled as a thin shell $S \in \RR^3$ of thickness 
$h > 0,$ parameterized over a domain $\Omega\subset\RR^2$ and an embedding $\phi: \Omega \times [-h/2,h/2] \rightarrow \mathbb{R}^3,$ with $S$ the image of $\phi.$ 
By the Kirchhoff-Love assumption, the entire shell volume can be represented
only in terms of the shell's mid-surface $r: \Omega \rightarrow \mathbb{R}^3,$
\begin{equation}
    {\phi}(u,v,t) = r(u,v)+t n(u,v),
\end{equation}
where $n=(r_u \times r_v)/||r_u \times r_v||$ 
is the midsurface normal, 
$r_u=\partial r/\partial u,\, r_v=\partial r/\partial v,$ 
and $t \in [-h/2,h/2].$  
The first and second fundamental forms of the surface are
\begin{gather} 
\label{eqn:fundamental} 
    a=F^T F, \quad
    b=-N^T F,\quad \textrm{where}\\
    F=[r_u\ r_v],\quad \textrm{and}\quad
    N=[n_u\ n_v].
\end{gather}
In this paper, we use $\tilde{a}$ 
and $\tilde{b}$ to denote the first and 
second fundamental forms in the initial rest state.
We assume that the parameterization is non-degenerate, i.e.,
matrix $F$ is rank 2. 

\paragraph{Plasticity}

\begin{figure*}[!ht]
            \includegraphics[width=0.9\hsize]{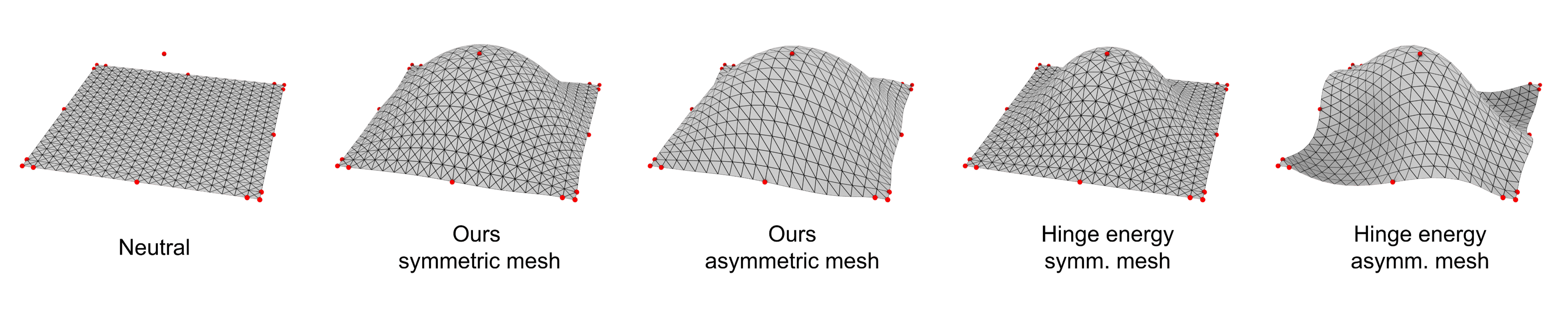}
            \vspace{-0.75cm}
        \caption[Comparison of our shape modeling method
        using a hinge-based elastic thin shell energy vs 
        energy defined via surface fundamental forms (as used in our method).]
        {\textbf{Comparison of our shape modeling method
        using a hinge-based elastic thin shell energy~\cite{Grinspun:2003:DS} vs 
        energy defined via surface fundamental forms~\cite{Chen:2018:PSO} 
        (as used in our method):}
        On asymmetric meshes, the hinge-based energy produces 
        substantial artefacts. The landmarks are shown in red.
        }
        \label{fig:plane}
\end{figure*}

We consider that a surface changes its shape through ``plasticity''
when it is assigned different first and second fundamental forms
$\bar{a}$ and $\bar{b}$ (at each location on the surface) from the initial rest state (Figure \ref{fig:plastic_schematic}).
Chen et al.~\shortcite{Chen:2018:PSO} employed plasticity for thin-shell forward simulation purposes, 
by directly modifying $\tilde{a}$ and $\tilde{b}$
into $\bar{a}$ and $\bar{b},$ using some procedural formulas. 
However, such an approach runs into a substantial limitation that
was readily apparent in our shape deformation system: 
arbitrary $\bar{a}$ and $\bar{b}$ may not satisfy the local compatibility conditions (Gauss-Codazzi equations~\cite{Weischedel:2012:ADG}).
In other words, given arbitrary first and second fundamental forms $\bar{a}$ and $\bar{b}$,
we can't guarantee that such a surface exists, not even locally.
According to our experiments, setting $\bar{a}$ and $\bar{b}$
without regarding to compatibility leads to poor search directions
and inability to decrease the energy for solving shape deformation (Section~\ref{sec:defo}).
Therefore, what is needed is a new approach that modifies 
the first and second fundamental forms in a compatible manner.
For each parameter $\sigma\in\Omega$ defining 
a surface point $r(\sigma)$, the vectors $t_1(\sigma) = r_u,\, t_2(\sigma)=r_v$ span the tangent plane. The unit normal at $r(\sigma)$ can be computed 
as $n=(t_1 \times t_2)/||t_1 \times t_2||$. 
Then, we can define a non-degenerate local frame at every point $\sigma$ by 
\begin{equation}
    D = \begin{bmatrix}t_1 & t_2 & n  \end{bmatrix}.
\end{equation}
Our idea is to modify the local frame at every point using a $3\times 3$ rotation matrix $R(\sigma)$ 
and a $2\times 2$ symmetric matrix $S(\sigma),$ leading to 
\begin{equation} \label{eqn:rs}
    D'=R \begin{bmatrix}t_1 & t_2 & n  \end{bmatrix} \begin{pmatrix} S & 0 \\ 0 & 1 \end{pmatrix}.
\end{equation}
Intuitively, we (anisotropically) scale the UV space at point $\sigma$ using $S$ and then change the local curvature by a spatially varying rotation matrix $R$. 
Because $(Rt_1) \times (Rt_2) = R(t_1 \times t_2),$ the normal in
the new frame $D'$ will always automatically be perpendicular to the new tangent plane, which is spanned 
by the columns of matrix $R[t_1 \,\, t_2] S.$ 
The new surface can be locally parameterized
via $\zeta \mapsto r'(\zeta) =  
R(\sigma) r(\sigma + s\,\, \zeta);$  
By derivation, one can verify that
\begin{equation}
\begin{aligned}
    &F' = [r'_u\quad r'_v] =R F S,\qquad
    &n'=Rn,
\end{aligned}
\end{equation}
which means that $D'$ is the local frame of the new embedding $r'.$
Therefore, the first and second fundamental 
forms $\bar{a}$ and
$\bar{b}$ derived from $r'$ are automatically compatible,
since they correspond to an actual locally defined surface.

Note that $S$ is an intrinsic variable which is unrelated to the embedding of the surface in $\RR^3,$ while $R$ is extrinsic, defining the embedding in $\RR^3.$ 
Our method cleanly decouples the intrinsic and extrinsic DoFs, 
so that $S$ and $R$ are independent of each other.
Observe that $R$ has 3 DoFs because it is a rotation matrix in 3D, 
and $S$ also has 3 DoFs because it is a symmetric matrix;
and therefore the space of local surface modifications is 6-dimensional.
In our implementation, we use the exponential map and the Rodrigues'
rotation formula to parameterize $R$ into a 3-dimensional vector $\theta$.
We also represent the symmetric matrix $S$ as a 3D vector $s$.

\paragraph{Difference to Polar Decomposition} 
We note that there is an alternative approach to change the local frame. 
Namely, one can perform polar decomposition of $D$ directly by $D=U A,$ 
where $U$ is a $3\times3$ rotation matrix and $A$ is a $3\times 3$ symmetric matrix~\cite{Kircher:2008:FFM}. 
The new frame then becomes $D'=U'A'$. 
Such an approach is less elegant for surface modeling 
because the change in the UV space follows the transformation $T=A'A^{-1}$, 
which is not a symmetric matrix, i.e., not a pure scaling matrix. 
Therefore, the transformation $T$ includes a rotational component in the UV space, 
which should enter into $U'$. Thus, this decomposition can't decouple the DoFs cleanly as we expected. 
Also, according to our experiments, 
one cannot obtain an effective search direction with this method to decrease the objective.

\paragraph{Elastic Energy} The second component of our modeling system
is elastic energy. 
Similarly to~\cite{Chen:2018:PSO}, for simplicity, we assume that the shell's material is homogeneous and isotropic, and adopt the St. Venant-Kirchhoff thin-shell 
constitutive law. We note that most of the in-plane deformation
is already handled by the plasticity in our method, and therefore the elasticity
only undergoes small in-plane strain, justifying the choice of St. Venant-Kirchhoff.
The elastic energy density can then be approximated (\cite{Weischedel:2012:ADG},~\cite{Chen:2018:PSO}) up to $O(h^4)$ by 
\begin{equation} \label{eqn:elastic_energy}
    W(u,v)=(\frac{h}{4}||\bar{a}^{-1}a-I||_{SV}^2 +\frac{h^3}{12}||\bar{a}^{-1}(b-\bar{b})||_{SV}^2)\sqrt{\det \bar{a}}
\end{equation}
where $||\cdot||_{SV}$ is the "St. Venant-Kirchhoff norm"
\begin{equation}
    ||M||=\frac{c_1}{2} \text{tr}^2 M +c_2 \text{tr}(M^2),
\end{equation}
and $c_1,c_2$ are material parameters related to Young's modulus $E$ and Poisson's ratio $v,$
\begin{equation}
    c_1=\frac{Ev}{1-v^2}, \quad c_2=\frac{E}{2(1+v)}.
\end{equation}
The elastic energy of the entire surface is
\begin{equation}
    V[r]=\int_{\Omega} W(u,v) du dv.
\end{equation}
Note that we first attempted to use the hinge-based energy to define
the bending energy~\cite{Grinspun:2003:DS},
but the results contained substantial artifacts and the resulting deformation was
not invariant of discretization, as shown in Figure~\ref{fig:plane}.  
Using the energy defined based on fundamental form energy produced no such bias~\cite{Chen:2018:PSO}.
Therefore, we abandoned the hinge-based energy.

\subsection{Discretization}

We approximate the mid-surface $r$ with a triangle mesh with $n$ vertices and $m$ triangles.
In the following paragraphs, we use \textbf{bold} symbol to represent the quantities for the entire mesh
and non-bold text to represent the quantities for a single vertex or element. 
The positions of the vertices $\boldsymbol{x}=[x_1,\ldots, x_n]\in\RR^{3n}$ 
take the place of the embedding function $r$. 
We assume that the first and second fundamental forms 
are constant over each triangle. 
We can now give a discrete form for Equations~\ref{eqn:fundamental},\ref{eqn:elastic_energy}.

\paragraph{Discrete Shell Model}

Let $f_{ijk}$ be a triangle with vertices $x_i,x_j,x_k,$ and let $\mathcal{T}$ be the canonical unit triangle with vertices $(0,0),(1,0),(0,1)$. Then, locally the face $f_{ijk}$ is embedded by the affine function
\begin{equation}
    r_{ijk}: \mathcal{T} \rightarrow \mathbb{R}^{3}, \quad r_{ijk}(u,v)=x_i +u (x_j-x_i) +v (x_k-x_i).
\end{equation}
The Euclidean metric on $f_{ijk}$ pulls back to the first fundamental form 
\begin{equation}
    a_{ijk}=\begin{bmatrix}
    ||x_j-x_i||^2  & (x_j-x_i) \cdot (x_k-x_i) \\
    (x_j-x_i) \cdot (x_k-x_i) & ||x_k-x_i||^2
    \end{bmatrix}.
\end{equation}
The $2\times 2$ matrix $a_{ijk}$ is always positive semi-definite. 
The second fundamental form can be 
discretized as
\begin{equation}
    b_{ijk}=2\begin{bmatrix}
    ({n}_i - {n}_j) \cdot (x_i-x_j)  & ({n}_i - {n}_j) \cdot (x_i-x_k)  \\
    ({n}_i - {n}_k) \cdot (x_i-x_j)  & ({n}_i - {n}_k) \cdot (x_i-x_k)
    \end{bmatrix},
\end{equation}
where ${n}_i$ is the mid-edge normal on the edge $e_i$ opposite to vertex $i$~\cite{Chen:2018:PSO}. If $e_i$ is a boundary edge, ${n}_i$ equals to the face normal, otherwise, ${n}_i$ is the average of the normals of the two faces incident at $e_i.$
Matrix $b_{ijk}$ is always symmetric but not in general positive-definite. 
We can now give a discrete formulation of the elastic energy 
\begin{equation}
    \begin{aligned}
    W_{ijk}=
    \bigl(\frac{h}{8}||\bar{a}_{ijk}^{-1}a_{ijk}-I||_{SV}^2 +\frac{h^3}{24}||\bar{a}_{ijk}^{-1}(\bar{b}_{ijk}-b_{ijk})||_{SV}^2\bigr)
    \sqrt{\det\bar{a}_{ijk}},
\end{aligned}
\end{equation}
where an additional division by two is due to the canonical triangle $\mathcal{T}$ having area $\frac{1}{2}$.

\paragraph{Discrete Plasticity} Let $\tilde{\boldsymbol{x}}$ represent the mesh positions in the rest state, and suppose the mesh at some location locally undergoes
a change $(R,S)$ as described earlier.
Then, by Equation~\ref{eqn:rs}, the plastic first 
fundamental form of the \textit{deformed} surface corresponding to $(R,S)$, 
in the discrete setting, is 
\begin{equation}
     \bar{a}_{ijk}=S^T \begin{bmatrix}
    ||\tilde{x}_j-\tilde{x}_i||^2 & (\tilde{x}_j-\tilde{x}_i)\cdot(\tilde{x}_k-\tilde{x}_i)\\
    (\tilde{x}_j-\tilde{x}_i)\cdot(\tilde{x}_k-\tilde{x}_i) & ||\tilde{x}_k-\tilde{x}_i||^2
    \end{bmatrix} S.   
\end{equation}
The second fundamental form is
\begin{equation}
     \bar{b}_{ijk}=2 S^T\begin{bmatrix}(\bar{n}_i-\bar{n}_j)^T R(\tilde{x}_i-\tilde{x}_j) & (\bar{n}_i-\bar{n}_j)^T R(\tilde{x}_i-\tilde{x}_k) \\
    (\bar{n}_i-\bar{n}_k)^T R(\tilde{x}_i-\tilde{x}_j) & (\bar{n}_i-\bar{n}_k)^T R(\tilde{x}_i-\tilde{x}_k)
    \end{bmatrix} S,
\end{equation}
where $\bar{n}_i$ represents the modified mid-edge normal of the plastic configuration after spatial varying rotations have been applied to triangle $f_{ijk}$ and its neighboring triangles in the initial state. 

\section{Plastic-Elastic Surface Shape Deformation}
\label{sec:defo}

\begin{figure}[!t]
        \includegraphics[width=1.0\hsize]{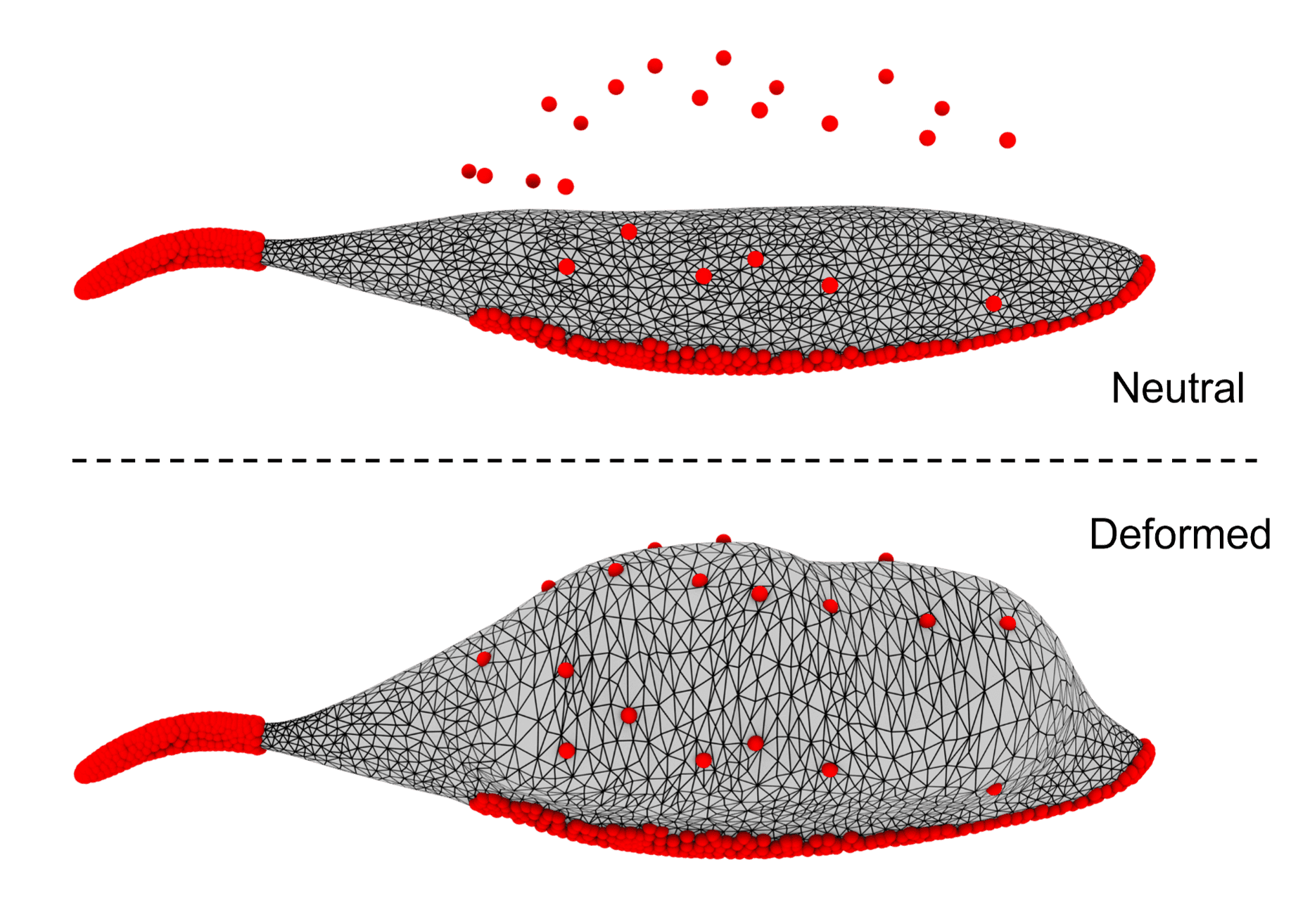}
        \vspace{-0.75cm}
        \caption[Deforming a muscle to match medical landmarks.]
        {\textbf{Deforming a muscle to match medical (MRI) landmarks:}
        The muscle is attached to the bone (not shown) on the lower side.
        We deform the template muscle to match the landmarks from a MRI scan,
        which causes large strains.
        The muscle MRI data was provided by the project~\cite{Wang:2021:MOP}.
        In contrast to~\cite{Wang:2021:MOP}, our method does not need a tetrahedral
        mesh and operates directly on the surface DOFs, whose number is substantially
        smaller than that of the volumetric mesh. We experimentally
        compared the performance: our method is $6\times$
        faster in the running time per iteration, and an overall $3\times$
        faster than~\cite{Wang:2021:MOP} to produce a result of equivalent quality.
        }
        \label{fig:muscle}
\end{figure}

Given an input triangle mesh, our goal is to deform the input mesh 
to match the user-defined target positions (landmarks), 
while ensuring smooth deformation (Figure~\ref{fig:muscle}). 
The position-based landmark constraints are freely defined by the user, 
whereby a selected point on an initial surface is manually corresponded to a target location. 
As explained in Section~\ref{sec:shell}, to model plasticity, 
we define a $3\times 3$ rotation matrix and a $2\times 2$ symmetric scaling matrix for each triangle.
After the parameterization, each triangle contains 6 DoFs given by $\theta,s$, respectively.
Finally, we group the DoFs of all triangles into a global vector $\boldsymbol{p} \in \mathbb{R}^{6m}$.

\subsection{Computing Vertex Positions from Plastic Strains}
Combining $\boldsymbol{p}$ with initial status $\tilde{\boldsymbol{a}},\tilde{\boldsymbol{b}}$
produces the current modified plasticity $\bar{\boldsymbol{a}},\bar{\boldsymbol{b}}$.
Then, minimizing the elastic energy corresponding to $\bar{\boldsymbol{a}},\bar{\boldsymbol{b}}$ 
gives the vertex positions $\boldsymbol{x}$.
To uniquely determine the local minimum, we define a proper boundary condition to eliminate the singularity of the elastic energy caused 
by its invariance on global rotation and translation.
To achieve this, we freeze a few selected vertices, 
which is commonly done in the existing methods.
In our implementation, if the landmarks already include 
anchored (i.e., permanently fixed) vertices, 
we simply select those vertices and treat them as fixed constraints.
If the landmarks don't include any anchored vertices, 
we fix $3$ arbitrary vertices which are not intended to move during the optimization. 
Therefore, the construction of $\boldsymbol{x}$ can be formulated as 
\begin{equation} \label{eqn:constraint_min}
    \argmin\limits_{\boldsymbol{x}} \mathcal{E}_e(\boldsymbol{x},\boldsymbol{p}) + \mathcal{E}_c(\boldsymbol{x})
\end{equation}
where $\mathcal{E}_e=\sum\limits_{ijk} W_{ijk}$ is the total discrete elastic energy, $\mathcal{E}_c=||\boldsymbol{C}\boldsymbol{x}-\boldsymbol{d}||^2$ is the quadratic position-based 
constraint energy, $\boldsymbol{C}$ is a constant selection matrix and $\boldsymbol{d}$ is the fixed position vector. 
Minimizing Equation~\ref{eqn:constraint_min} is equivalent to solving for 
the stationary point of:
\begin{equation}
    \boldsymbol{f}_e(\boldsymbol{x},\boldsymbol{p})+\boldsymbol{f}_c(\boldsymbol{x}) =0,
\end{equation}
where $\boldsymbol{f}_e=\frac{\partial \mathcal{E}_e}{\partial \boldsymbol{x}}$ is 
elastic force, and $\boldsymbol{f}_c=\frac{\partial \mathcal{E}_c}{\partial \boldsymbol{x}}$ is constraint force.

\subsection{Shape Deformation Using Plastic Strains}

\begin{figure}[!ht]
    \centering
    \includegraphics[width=1.0\hsize]{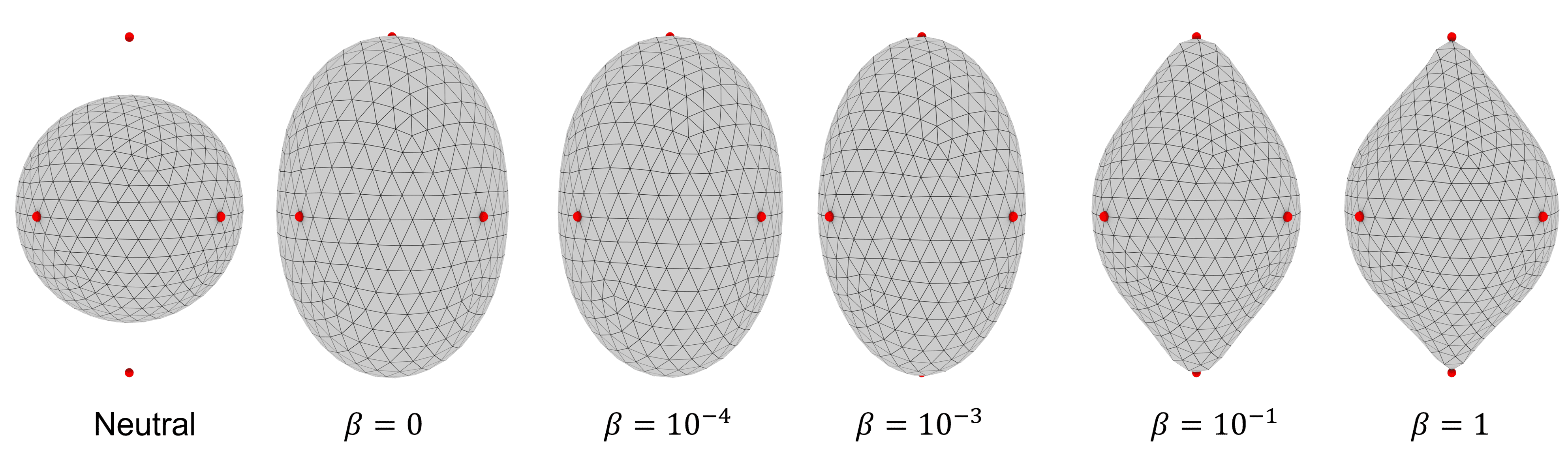}
    \vspace{-0.75cm}
    \caption{Exploring parameter $\beta$ in Equation~\ref{eqn:optimization}.
    Parameter $\beta$ permits us to control the tradeoff between
    smoothness of the applied deformations (low $\beta$) vs staying similar 
    to the input shape (high $\beta$). [Wang et al. 2012] and other
    elastic-energy-based methods correspond to choosing a 
    high value of $\beta,$ hence they produce spikes when the landmarks
    dictate large strains.
    }
    \label{fig:shphere}
\end{figure}

We would like to discover vertex positions $\boldsymbol{x}$
and a smooth change of plastic strains $\boldsymbol{p}$ 
such that the surface meets user-defined constraints:
\begin{gather} 
\label{eqn:optimization}
    \argmin\limits_{\boldsymbol{p},\boldsymbol{x}} ||\boldsymbol{L}\boldsymbol{p}||^2 +\alpha  \mathcal{E}_{l}(\boldsymbol{x}) +\beta ||\boldsymbol{p}-\boldsymbol{p}_0||^2     \\
    \text{s.t. } \boldsymbol{f}_e(\boldsymbol{p},\boldsymbol{x}) +\boldsymbol{f}_{c}(\boldsymbol{x})=0,
\end{gather}
where $\boldsymbol{L}$ is the Laplacian operator on the surface mesh,
and $\boldsymbol{p}_0$ is the initial value of $\boldsymbol{p}$, i.e., identity matrices for $\boldsymbol{s}$ and zeros for $\boldsymbol{\theta}$.
The Laplacian term enforces the smoothness of $\boldsymbol{p}$, i.e.,
the change of plasticity.
Term $||\boldsymbol{p}-\boldsymbol{p}_0||^2$ controls the smoothness of the deformation,
as depicted in Figure~\ref{fig:shphere}.
The terms $\mathcal{E}_{l}(\boldsymbol{x})$ is the quadratic landmark energy 
similar to the constraint energy, and $\alpha\geq 0$ is the landmark weight. 

\subsection{Plastic Strain Laplacian}
\label{sec:laplacian}
We now explain how we compute the Laplacian $\boldsymbol{L}$.
We compute different Laplacian matrices for $\boldsymbol{\theta}$ and $\boldsymbol{s}$ in $\boldsymbol{p}$.
Suppose that there are two neighboring triangles $i$ and $j$ on the surface mesh. 
The technical difficulty in computing the Laplacian of $\boldsymbol{s}$ is that 
$\boldsymbol{s}$ is defined in the local coordinate system of each triangle.
The coordinate systems of triangles $i$ and $j$ 
(denoted as ${\Omega}_i$ for triangle $i$ and ${\Omega}_j$ 
for triangle $j$) are different and are arbitrarily
rotated, hence we cannot directly apply the Laplacian to entries of $\boldsymbol{s}.$
Therefore, we calculate the rotation matrix $Q_{ji}$ that 
transforms material coordinates from ${\Omega}_j$ to ${\Omega}_i,$
and use it to transform $S_j$ into the coordinate system of triangle $i.$
In this manner, we consistently define the Laplacian operator at triangle $i$ as
\begin{equation}
    L_{s,i}=\sum\limits_{j \in \mathcal{N}(i)} \textrm{mat}(s_i)-Q_{ji} \textrm{mat}(s_j) Q_{ji}^{T},
\end{equation}
where $\mathcal{N}(i)$ represents the triangles adjacent to triangle $i$
and $\textrm{mat}(s)$ converts the vector $s$ back to its matrix form.

The rotation matrix $R$ is always given in the global coordinate system,
and hence no special treatment is required; we directly apply the 
triangle-wise Laplacian:
\begin{equation}
    L_{\theta,i}=\sum\limits_{j \in \mathcal{N}(i)} \theta_i - \theta_j,
\end{equation}
where $\theta=\textrm{log}(R)\in\RR^3$ is the exponential map representation of the rotation $R.$ 
It should be noted that this measurement is not invariant of global rotation.
However, our rotation field is local with respect to the rest shape.
Due to performance considerations, 
we use this simple method and found that
it works well in all the benchmarks and test cases.If global rotation is needed, we can insert an additional first step, namely first performing rigid Iterative Closest Point (ICP) to align the orientation of the object. 

\subsection{Optimization}
To solve our optimization problem, we employ a Gauss-Newton solver 
in the same way in~\cite{Wang:2021:MOP}.
To improve the convergence of Gauss-Newton solver, 
we update the rest shape by the output vertex positions of the previous iteration at each Gauss-Newton iteration.
To compensate for this update, we also re-initialize $\boldsymbol{p}$ to $\boldsymbol{p}_0$ at the beginning of each iteration. In our implementation, we use inexact Hessian \cite{Chen:2018:PSO} and project local Hessian to the cone of symmetric positive semi-definite matrices prior to assembly to improve the performance.

\begin{figure*}[!ht]
            \includegraphics[width=1.0\hsize]{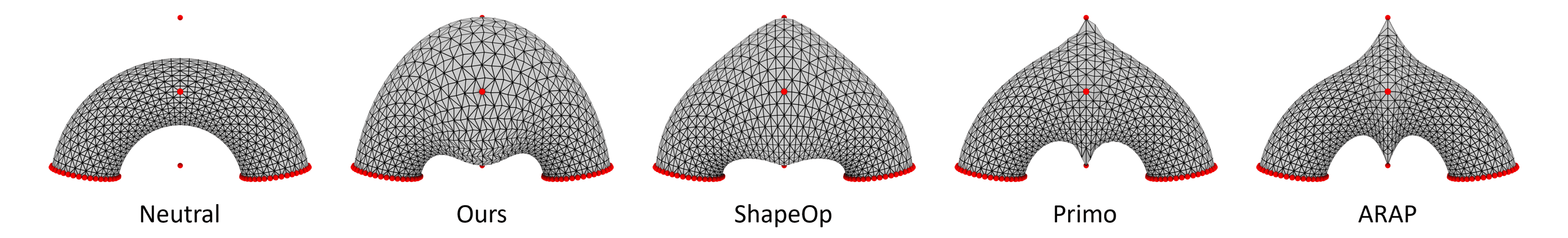}
            \vspace{-0.75cm}
        \caption[Large strain editing of a tube.]
        {\textbf{Large strain editing of a bending tube:}
        Similar to \cite{Wang:2021:MOP}, we use landmark (red dots) to grow a bending tube in the middle part. Our method can produce a symmetric and smooth "arch". The result of shapeOp is not as smooth as ours. The Primo and ARAP suffer from sharp spikes. We tuned the parameters of both ShapeOp and Primo and tried various settings and the results weren't improved.}
        \label{fig:arch}
\end{figure*}

\begin{figure}[!h]
            \includegraphics[width=1.0\hsize]{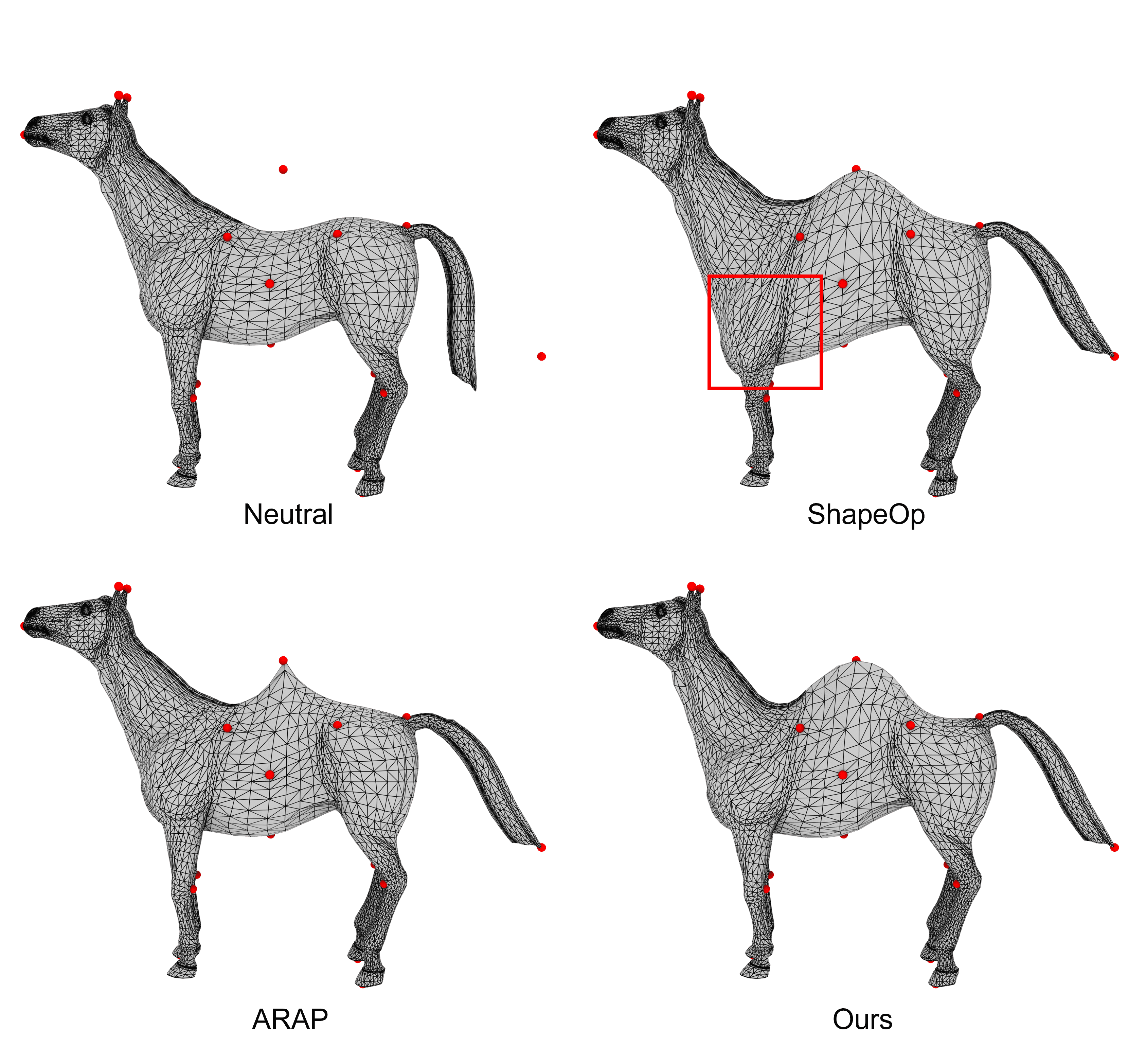}
            \vspace{-0.75cm}
        \caption[Comparison to ARAP and ShapeOp]
        {\textbf{Comparison to ARAP~\cite{Sorkine:2007:ARA} and ShapeOp~\cite{Bouaziz:2014:PD}:}
        We apply a single landmark (red dot) to add a ``hump'' onto the horse and another landmark to lift the tail slightly,
        while preserving the remaining shape.
        Our method produces a smooth large deformation.
        ARAP produces a spike, and ShapeOp distorts a hoof, generates a strange stretched belly
        and produces a non-smooth hump. We tweaked
        the ShapeOp parameters and tried multiple settings; 
        a better result could not be produced.
        }
        \label{fig:horse}
\end{figure}

\begin{figure}[!h]
    \centering
    \includegraphics[width=1.0\hsize]{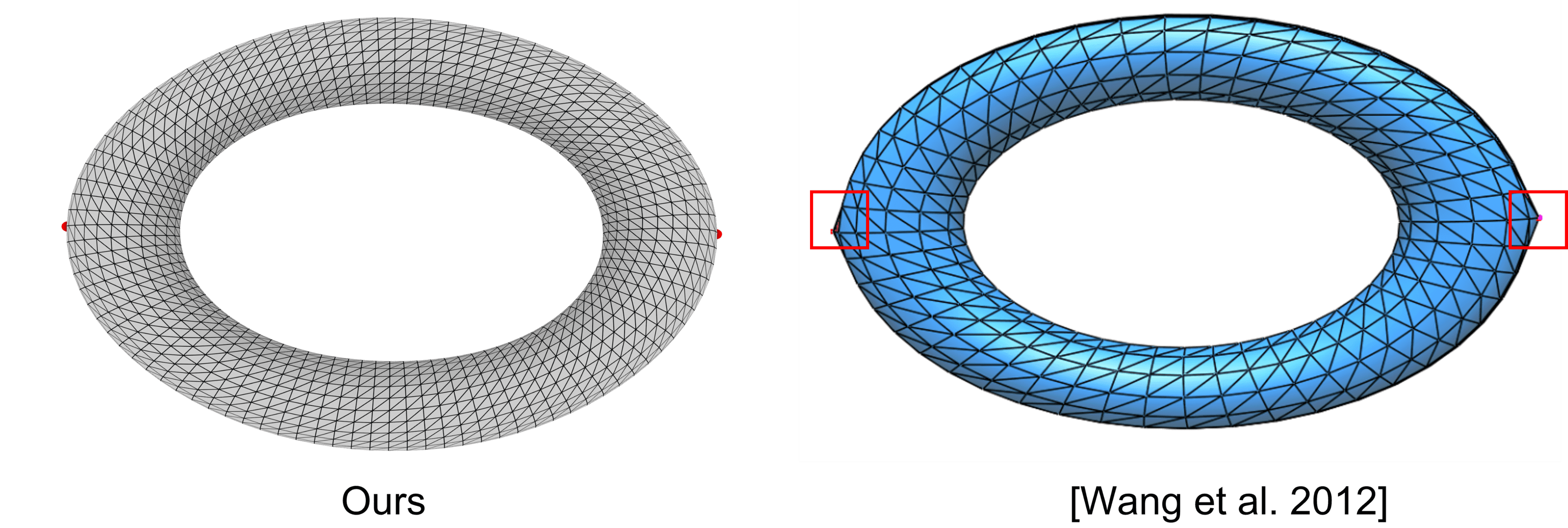}
    \vspace{-0.75cm}
    \caption{\textbf{Comparison to ~\shortcite{wang2012linear}}
    Wang et al.~\shortcite{wang2012linear} does not address large strain editing;
    it produces spikes at the constraints, as shown in their result.
    Our method produces smooth large deformations.
    \label{fig:torus}}
\end{figure}

\begin{figure}[!h]
    \centering
    \includegraphics[width=0.8\hsize]{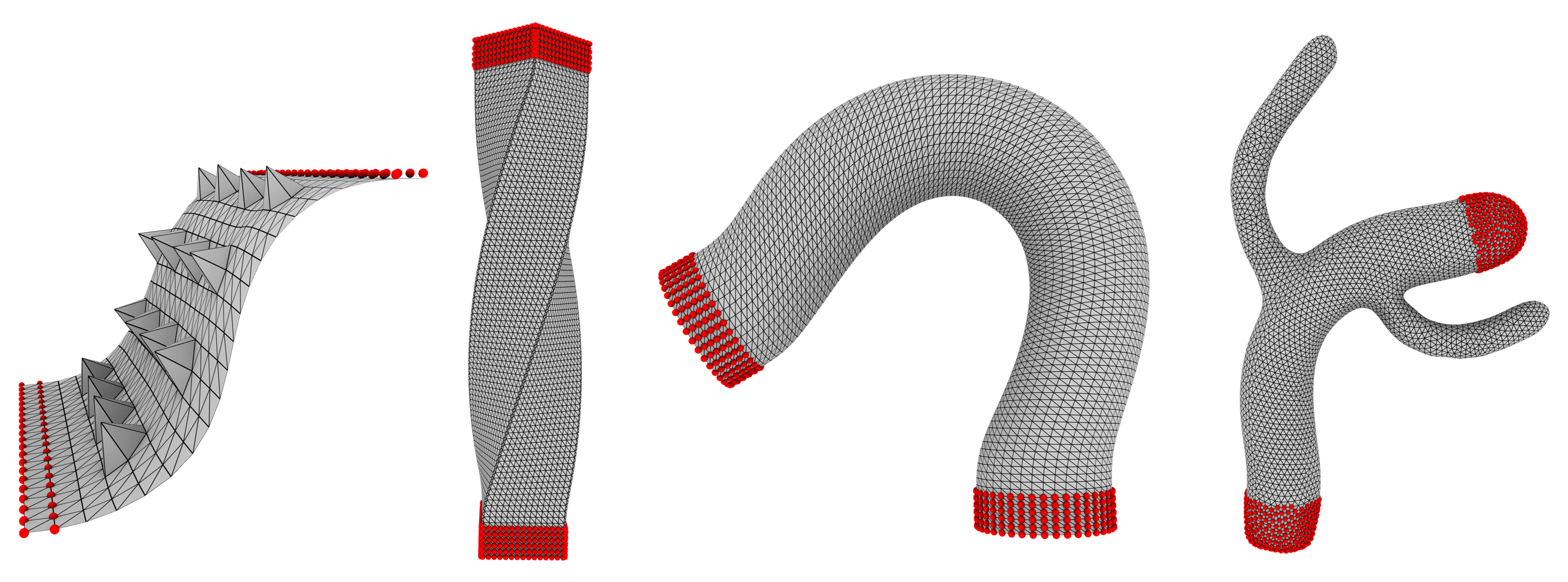}
    \vspace{-0.25cm}
    \caption{Our method passes the standard shape deformation benchmark~\cite{Botsch:2008:OLV}.}
    \label{fig:benchmark}
\end{figure}

\section{Results}
\label{sec:results}

\begin{figure}[!t]
        \includegraphics[width=1.0\hsize]{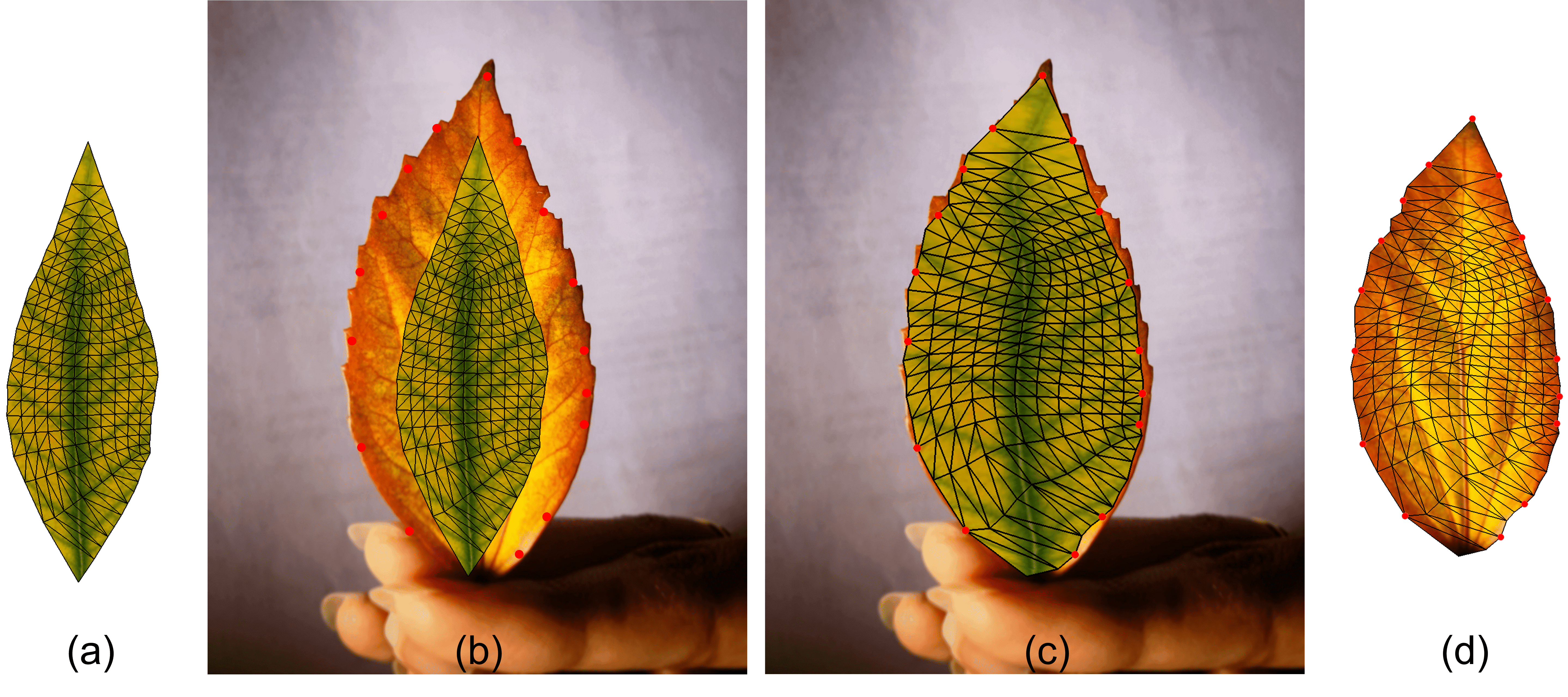}
        \vspace{-0.5cm}
        \caption[Shape deformation with large strains.]
        {\textbf{Shape deformation with large strains.}
        We reshaped the input leaf mesh (a) to match the leaf landmarks 
        that we manually determined based on a photograph of the target leaf
        (seen in (b); the hand holding the brown leaf is a photograph).
        Part (b) shows the target landmarks
        that we determined manually based on the photograph in red.
        This produces a leaf mesh (c) whose shape matches the photograph and whose
        mesh topology matches the input leaf. Such mesh correspondence can be used
        for further mesh processing applications 
        (e.g., morphing, texture transfer, procedural leaf generation).
        In (d), we transfer the texture from the photograph onto our deformed
        leaf mesh.
        }
        \label{fig:leaf1}
\end{figure}

We give several examples that demonstrate shape deformation involving
large rotations and strains while meeting non-trivial user constraints
(Figures~\ref{fig:arch},\ref{fig:horse},\ref{fig:bunny},\ref{fig:torus}).
These examples also provide comparisons to 
ARAP~\cite{Sorkine:2007:ARA}, Primo~\cite{Botsch:2006:PCP}, 
\cite{wang2012linear}, and ShapeOp~\cite{Bouaziz:2014:PD}.
We also test our model using standard shape deformation benchmark~\cite{Botsch:2008:OLV}, 
as shown in Figure~\ref{fig:benchmark}.Figure ~\ref{fig:leaf1} gives another interesting application.
Table~\ref{tab:perf} gives the performance of each demo.

\begin{table}[!htp]\centering
\caption{\textbf{The setup and performance of each example.}
The first half of the table shows the number of vertices (\#vtx), the number of triangles (\#tri),
and the number of landmarks (\#land).
In addition, the table also lists the performance of each example.
Column \#iter denotes the number of Gauss-Newton iterations performed in each example.
Column $\epsilon_{\textrm{land}}$ gives
the errors of landmark constraints,
given that the diagonal length of the bounding box of the mesh is 100 for all examples.
Finally, column $t_{\textrm{opt}}$ 
shows the optimization time cost per iteration.
We also measured the time cost per iteration of the volumetric method from~\cite{Wang:2021:MOP} for
the muscle example. For muscle, it takes 59s per iterations and 4 iterations in total,
6x slower per iteration and 3x slower in total than our method.}
\label{tab:perf}
\scriptsize
\begin{tabular}{l|rrrrrrrrr}\toprule
&\#vtx      &\#tri   &\#land  &\#iter& $t_{\textrm{opt}}$ & $\epsilon_{\textrm{land}}$ \\\midrule
leaf        & 233    & 411    & 16   & 13 & 0.3s  &0.06 \\
plane       & 441    &    800 &  17  &  6 &  0.5s &0.15  \\
squarespike & 441    & 800    & 84   & 13 & 0.7s &0.35 \\
sphere      & 642    &1,280   &6     &5   &0.6s   &0.57  \\
arch        &  1,756 &  3,508 & 200  & 13 &3.6s  &0.22  \\
torus       & 2,304  & 4,608  & 50   & 5  & 2.4s  & 0.11 \\
bunny       &2,503   &  4,968 &   9  & 11 &  3.4s &0.19  \\
muscle      &  3,288 &  6,572 & 650  &  5 & 9.2s &0.21  \\
cylinder    & 4,802  & 9,600  & 482  & 20 & 11.6s & 0.01 \\
cactus      & 5,261  & 10,518 & 864  & 14 & 15.5s & 0.05 \\
twist bar   & 6,084  & 12,106 & 962  & 15 & 20.7s & 0.15 \\
horse       &8,431   & 16,843 & 21   & 9  & 8.5s  & 0.16 \\
fish        & 10,786 & 21,568 &  26  & 18 & 11.1s &0.14  \\
\bottomrule
\end{tabular}
\end{table}

\section{Conclusion}

We gave a new differential representation for surface
deformation that enables modeling shape deformations
consisting of large spatially varying rotations and strains.
Our formulation could be extended to other manifolds
embedded in higher-dimensional spaces, for example,
curves in three dimensions; or, more generally,
$m$-dimensional manifolds embedded in $\RR^n,$
for $m<n.$ If local rotations across the surface
accumulate beyond $2\pi,$ the spatially varying
rotation field may have a discontinuity where the 
rotation ``jumps'' by a multiple of $2\pi.$ 
While we did not run into this issue in our examples,
this is a well-known
problem in modeling rotational fields on surfaces, and
there are standard solutions for it.
While our method produces quality shapes that contain
large strains and rotations and that precisely meet
user constraints, the price to pay for this is
that the method runs offline and is not interactive.
This is because our method needs to solve an optimization problem
for the plastic strains, which necessitates solving
large sparse systems of equations and evaluating
many energy, force and Hessian terms. The predominant bottleneck
of our method is the solution of large sparse linear systems
during the optimization process. We already used 
inexact Hessians~\cite{Chen:2018:PSO} to speed up these solves;
further speedups could be obtained using multigrid or specially
designed preconditioners for our problem.

\begin{acks}
This research was sponsored in part by
NSF (IIS-1911224), USC Annenberg Fellowship
to Jiahao Wen and Bohan Wang, Bosch Research and Adobe Research.
\end{acks}

\bibliographystyle{ACM-Reference-Format}
\bibliography{plasticShell.bib}



%
%

\end{document}